\documentclass[twocolumn,showpacs,prl,amsmath,amssymb,floatfix]{revtex4}
\usepackage{amsmath,amssymb}
\usepackage{bm}
\usepackage{epsfig}

\newcommand{\bd}{\bm}

\begin{document}
  \title{Persistent spin currents in mesoscopic Heisenberg rings}
  
  \author{Florian Sch\"{u}tz, Marcus Kollar, and Peter Kopietz}
  
  \affiliation{Institut f\"{u}r Theoretische Physik, Universit\"{a}t
    Frankfurt, Robert-Mayer-Strasse 8, 60054 Frankfurt, Germany}

  \date{July 7, 2003}

  \begin{abstract}
    We show that at low temperatures $T$ an inhomogeneous radial
    magnetic field with magnitude $B$ gives rise to a persistent
    magnetization current around a mesoscopic ferromagnetic Heisenberg
    ring.  Under optimal conditions this spin current can be as large
    as $ g \mu_{\text{B}} (T / \hbar) \exp [ - 2 \pi ( g
    \mu_{\text{B}} B / \Delta )^{1/2} ]$, as obtained from
    leading-order spin-wave theory.  Here $g$ is the gyromagnetic
    factor, $\mu_{\text{B}}$ is the Bohr magneton, and $\Delta$ is the
    energy gap between the ground state and the first spin-wave
    excitation.  The magnetization current endows the ring with an
    electric dipole moment.
  \end{abstract}




  \maketitle
  
  The controlled fabrication of submicron devices has opened the door
  to a rich new field of theoretical and experimental physics.  At low
  temperatures these devices are mesoscopic in the sense that their
  quantum states must be described by coherent wave functions extending
  over the entire system. Then the usual assumptions underlying the
  averaging procedure in statistical mechanics are not necessarily
  valid, and quantum-mechanical interference effects become important
  \cite{Imry97}.
  
  A prominent example is persistent currents in mesoscopic normal
  metal rings threaded by a magnetic flux \cite{Imry97}.  Although
  this phenomenon was predicted long ago \cite{Hund38,Buttiker83}, the
  experimental difficulties in measuring persistent currents in an
  Aharonov-Bohm geometry were only overcome in the past decade
  \cite{Levy90,Chandrasekhar91,Mailly93}.  Surprisingly, for metallic
  rings in the diffusive regime the observed currents were much larger
  than predicted by theory \cite{Imry97}.  On the other hand, in the
  ballistic regime \cite{Mailly93} the order of magnitude of the
  observed current can be explained with a simple model of free
  fermions moving on a
  ring pierced by a magnetic flux $\phi$. Then the stationary energies
  are $\epsilon_n = \hbar^2 k_n^2 / 2 m_{\ast}$, where $k_n = \frac{ 2
    \pi }{L} ( n - \frac{ \phi}{ \phi_0} )$, $n = 0, \pm 1, \pm 2 ,
  \ldots$, are the allowed wavevectors for a ring with circumference
  $L$.  Here $\phi_0 = hc/e$ is the flux quantum and $m_{\ast}$ is the
  effective mass of the electrons.  In the simplest approximation, one
  may calculate the current $I = - c \partial \Omega_{\rm{gc}} ( \phi
  ) / { \partial \phi}$ at constant chemical potential $\mu$ from the
  flux-dependent part of the grand canonical potential
  $\Omega_{\rm{gc}} ( \phi )$.  At finite temperature $T$, one obtains
  for spinless fermions \cite{Cheung88}
  \begin{equation}
    {I} = 
    \frac{-e}{ L} \sum_{n}
    \frac{v_n}{e^{ ( \epsilon_n - \mu  )/ T } + 1 }
    \,,
    \label{eq:Icharge}
  \end{equation} 
  where $v_n = \hbar k_n / m_{\ast}$. For $T \ll \mu$ the
  amplitude of the current is $I_{\rm{max}} \approx -e v_F / L$ (where
  $v_F$ is the Fermi velocity), in agreement with experiment
  \cite{Mailly93}.
  
  In this Letter, we show that Heisenberg spin chains in inhomogeneous
  magnetic fields can be used to realize a spin current analogue of
  mesoscopic persistent currents in normal metal rings. Note that in the 
  presence of spin-orbit coupling spin currents in spin chains can also 
  be driven by inhomogeneous electric fields \cite{Cao97}, due to the
  Aharonov-Casher effect \cite{Aharonov84}.  As detailed later on, the
  magnetization current is carried by magnons and endows the ring with
  an {\it{electric}} dipole field, which is the counterpart of the
  magnetic dipole field associated with the persistent charge current
  in a normal metal ring.  We find that for realistic parameters the
  spin analogues of the experiments in
  Refs.~\onlinecite{Levy90,Chandrasekhar91,Mailly93} require the
  detection of a potential drop on the order of nanovolts.
  
  Due to its relevance for information processing based on spin
  degrees of freedom, the problem of magnetization transport has
  received a lot of attention recently \cite{Awschalom02}, especially
  for systems where spin currents are carried by itinerant electrons
  \cite{Loss90,Stern92,Gao93,Konig01,Tatara02,Malshukov02,Shen02}.
  Magnetic insulators also show interesting spin transport phenomena
  \cite{Gorelik02,Meier02}.  Very recently, Meier and Loss
  \cite{Meier02} calculated the mesoscopic spin conductance for
  Heisenberg-type systems in a two-terminal geometry.  Here we
  consider the same problem in a ring geometry and for inhomogeneous
  magnetic fields that lead to noncoplanar spin configurations. It is
  known that such configurations can lead to dissipationless transport
  of charge and spin \cite{Loss90,Konig01,Tatara02}.  In the following,
  we use linear spin-wave theory to derive the mesoscopic persistent
  spin current circulating in the ring, corresponding to an
  {\it{infinite}} spin conductance, and explicitly obtain the
  associated electric field.

\begin{figure}[tb]
    \begin{center}
      \epsfig{file=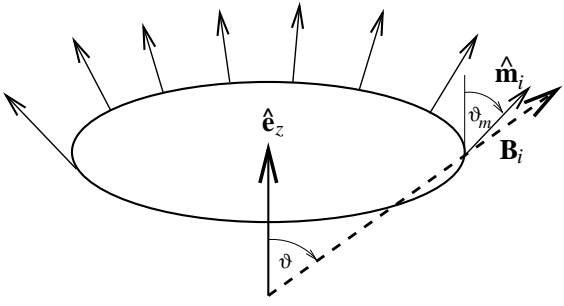,width=75mm}
    \end{center}
    \vspace{-4mm}
    \caption{%
      Classical spin configuration $\hat{\bd{m}}_i$ of a
      nearest-neighbor ferromagnetic Heisenberg ring in a radial
      magnetic field ${\bd{B}}_i$.}
    \label{fig:angle}
  \end{figure}

  Let us start with a general Heisenberg Hamiltonian
  \begin{equation}
    \hat{H} = \frac{1}{2} \sum_{  i,j}   J_{ij} {\bd{S}}_i \cdot
    {\bd{S}}_j   - g \mu_{\text{B}} \sum_{i  }  {\bd{B}}_i \cdot
    {\bd{S}}_i
    \,,
    \label{eq:Hamiltonian}
  \end{equation} 
  where the sums are over all sites ${\bd{r}}_i$ of a chain with
  periodic boundary conditions, $J_{ij}$ are general exchange
  couplings, and ${\bd{S}}_i$ are spin-$S$ operators normalized such
  that ${\bd{S}}_i^2 = S ( S+1 )$.  The last term in
  Eq.~(\ref{eq:Hamiltonian}) is the Zeeman energy associated with an
  inhomogeneous magnetic field ${\bd{B}}_i = {\bd{B}} ( {\bd{r}}_i )$.
  We assume that the magnetic field at each lattice site is
  sufficiently strong to induce permanent magnetic dipole moments
  ${\bd{m}}_i = g \mu_{\text{B}} \langle {\bd{S}}_i \rangle$, not
  necessarily parallel to $\bd{B}_i$, where $\langle \ldots \rangle$
  denotes the usual thermal average.  Moreover, we assume that the
  unit vectors $\hat{\bd{m}}_i$ $=$ ${\bd{m}}_i / | {\bd{m}}_i |$
  trace out a finite solid angle $\Omega$ on the unit sphere in
  order-parameter space as we move once around the chain.  The
  simplest geometry is a ferromagnetic ring in a crown-shaped magnetic
  field, as illustrated in Fig.~\ref{fig:angle}.
  
  In the general case, the Hamiltonian (\ref{eq:Hamiltonian}) implies 
  the equation of motion
  \begin{equation}
    \hbar \frac{ \partial   {\bd{S}}_i}{\partial t }    
    +   {\bd{h}}_i 
    \times  {\bd{S}}_i +
    \sum_{j } {\bd{I}}_{ i \rightarrow j}    
    = 0
    \,,
    \label{eq:eom}
  \end{equation}
  where ${\bd{h}}_i = g \mu_{\text{B}} {\bd{B}}_i$, and $ {\bd{I}}_{ i
    \rightarrow j} = {J_{ij}} {\bd{S}}_i \times {\bd{S}}_j$ is the
  spin current from site ${\bd{r}}_i$ to site ${\bd{r}}_j$.  From
  Eq.~(\ref{eq:eom}) it is easy to show that in equilibrium $ \sum_{j}
  \hat{\bd{m}}_i \cdot \langle {\bd{I}}_{ i \rightarrow j} \rangle =
  0$ which is Kirchhoff's law for spin currents
  \cite{footnotetensor}.
  
  It is convenient to decompose the spin operators as ${\bd{S}}_i =
  S^{\parallel}_i \hat{\bd{m}}_i + {\bd{S}}^{\bot}_i$, with
  ${\bd{S}}_i^{\bot} \cdot \hat{\bd{m}}_i = 0$.  Substituting this
  into Eq.~(\ref{eq:Hamiltonian}) we obtain $\hat{H} =
  \hat{H}^{\parallel} + \hat{H}^{\bot} + \hat{H}^{\prime}$, with
  \begin{eqnarray}
    \hat{H}^{\parallel} & = & \frac{1}{2} \sum_{  i,j}  J_{ij}  
    \hat{\bd{m}}_i \cdot \hat{\bd{m}}_j
    {{S}}_i^{ \parallel} 
    {{S}}_j^{\parallel}   - \sum_{i  }  {\bd{h}}_i  \cdot \hat{\bd{m}}_i 
    {{S}}_i^{\parallel}
    \,,
    \label{eq:Hparallel}
    \\
    \hat{H}^{\bot} & = & \frac{1}{2} \sum_{  i,j}  J_{ij}  
    {\bd{S}}^{\bot}_i \cdot {\bd{S}}^{\bot}_j  
    \,,
    \label{eq:Hbot}
    \\
    \hat{H}^{\prime} & = &  - \sum_{  i}   {\bd{S}}^{\bot}_i
    \cdot \bigl(  {\bd{h}}_i -
    \sum_j J_{ij}  
    {{S}}_j^{ \parallel}   \hat{\bd{m}}_j   \bigr)
    \,.
    \label{eq:Hrest}
  \end{eqnarray} 
  To develop the standard semiclassical spin-wave theory, we expand in
  the small parameter $1/S$. The leading-order term [i.e.,
  Eq.~(\ref{eq:Hparallel}) with $ S^{\parallel}_i$ replaced by $S$]
  yields the classical energy.  A necessary condition for its minimum
  is the invariance under small variations $\delta\hat{\bd{m}}_i$.
  This leads to the condition
  \begin{equation}
    \hat{\bd{m}}_i
    \times
    \Big(
    {\bd{h}}_i - S \sum_j J_{ij}  \hat{\bd{m}}_j
    \Big)
    = 0
    \,, 
    \label{eq:classical}
  \end{equation} 
  which shows that the magnetization aligns parallel to the sum of
  external and exchange field, as expected.  For given ${\bd{h}}_i $
  and $J_{ij}$, this is a system of non-linear equations for the spin
  directions $\hat{\bd{m}}_i$ in the classical ground state.  To study
  fluctuations, we expand the transverse components as $
  {\bd{S}}^{\bot}_i $ $=$ $\frac{1}{2} \sum_{ p = \pm } S_i^{-p}
  {\bd{e}}^{p}_i $, with spherical basis vectors ${\bd{e}}^{p}_i $ $=$
  $ \hat{\bd{e}}^1_i + i p \hat{\bd{e}}^2_i$, $p $ $=$ $ \pm$, where $
  \left\{ \hat{\bd{e}}_i^{1}, \hat{\bd{e}}_i^2 , \hat{\bd{m}}_i
  \right\}$ is a local orthogonal triad of unit vectors. The
  transverse part of our spin Hamiltonian can then be written as
  \begin{equation}
    \hat{H}^{\bot}  =  \frac{1}{8} \sum_{  i,j}  \sum_{ p, p^{\prime} } J_{ij}  
    ( {\bd{e}}^{p}_i \cdot {\bd{e}}^{p^{\prime}}_j )
    {{S}}^{-p}_i 
    {{S}}^{-p^{\prime}}_j  
    \,.
    \label{eq:Hbot2}
  \end{equation}
  For explicit calculations, we use the standard representation 
  $S^{\parallel}_i$ $=$ 
  $S-b_i^{\dagger}b_i^{\phantom{\dagger}}$ and $S^{+}_i$ $=$
  $(S^{-}_i)^\dagger$ $=$
  $\sqrt{2S}\,b_i^{\phantom{\dagger}}\,[1+O(S^{-1})]$ in terms of
  canonical boson operators $b_i$.
  Note that $\hat{H}^{\bot}$ $=$ $O(S)$, whereas $\hat{H}^{\prime}$
  $=$ $O(S^{1/2})$ due to Eq.~(\ref{eq:classical}).

\begin{figure}[tb]
    \begin{center}
      \epsfig{file=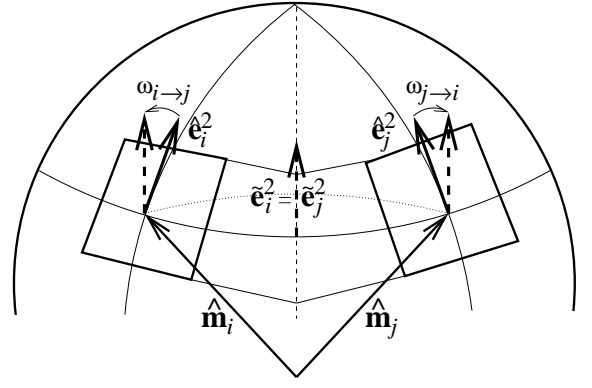,width=75mm}
    \end{center}
    \vspace{-4mm}
    \caption{%
      Definition of the rotated basis vectors and the corresponding 
      rotation angles $\omega_{ i
        \rightarrow j}$ and $\omega_{ j \rightarrow i}$.  The rotated
      vectors $\tilde{\bd{e}}_i^2 = \tilde{\bd{e}}_j^2$ lie in the
      intersection of the two tangent planes associated with
      $\hat{\bd{m}}_i$ and $\hat{\bd{m}}_j$, shown as a dashed line.      
      The dotted line is the geodesic connecting the two sites
      on the unit sphere in order-parameter space.}
    \label{fig:tangentplane}
  \end{figure}

  While the $\hat{\bd{m}}_i$ are fixed by
  Eq.~(\ref{eq:classical}), there is a remaining local $U(1) $
  gauge freedom associated with the rotation of the transverse basis
  vectors ${\bd{e}}^{p}_i$.  Let us rewrite the scalar product $
  {\bd{e}}^{p}_i \cdot {\bd{e}}^{p^{\prime}}_j$ such that the local
  gauge invariance is manifest.  Rotating the transverse basis vectors
  with an angle $\omega_{i \rightarrow j}$ around the local normal
  $\hat{\bd{m}}_i$ leads to ${\bd{e}}_i^{p} = e^{i p \omega_{i
      \rightarrow j} } {\tilde{\bd{e}}}_i^p$, as shown in
  Fig.~\ref{fig:tangentplane}.
  If we choose the rotated basis vectors $\tilde{\bd{e}}^p_i$ and
  $\tilde{\bd{e}}^p_j$ associated with two neighboring points
  $\hat{\bd{m}}_i$ and $\hat{\bd{m}}_j$ on the unit sphere such that
  $\tilde{\bd{e}}_i^{2}$ and $\tilde{\bd{e}}_j^2$ are equal and
  parallel to $\hat{\bd{m}}_i \times \hat{\bd{m}}_j$, 
  then $\tilde{{\bd{e}}}_i^p \cdot \tilde{\bd{e}}_j^{p^{\prime} } = (
  \hat{\bd{m}}_i \cdot \hat{\bd{m}}_j - p p^{\prime} )$.  Rotating
  back to the original basis, we arrive at
  \begin{equation}
    {\bd{e}}_i^p \cdot {\bd{e}}_j^{p^{\prime} } =
    ( \hat{\bd{m}}_i \cdot \hat{\bd{m}}_j  - p p^{\prime} )
    \exp[  
      i  p \omega_{ i \rightarrow  j } + i 
      p^{\prime} \omega_{ j \rightarrow i}]
    \,.
    \label{eq:omegaphase}
  \end{equation}  
  Note that the condition $\tilde{\bd{e}}_i^{2} = \tilde{\bd{e}}_j^2$
  implies ${\rm{Im}} [ \tilde{{\bd{e}}}_i^p \cdot
  \tilde{\bd{e}}_j^{p^{\prime} } ] = 0$.  Geometrically, this means
  that $\tilde{{\bd{e}}}_i^p$ and $\tilde{{\bd{e}}}_j^p$ are related
  by parallel transport \cite{Shapere89} along the shortest path (i.e.,
  a geodesic, corresponding to a rotation around an axis parallel to
  $\hat{\bd{m}}_i \times \hat{\bd{m}}_j$) connecting the points
  $\hat{\bd{m}}_i$ and $\hat{\bd{m}}_j$ on the surface of the unit
  sphere.  
  Substituting Eq.~(\ref{eq:omegaphase})
  into Eq.~(\ref{eq:Hbot2}), we obtain
  \begin{multline}
    \hat{H}^{\bot}  =  
    \frac{1}{8}
    \sum_{i,j} J_{ij}
    \Bigl[
    ( 1 + \hat{\bd{m}}_i \cdot \hat{\bd{m}}_j ) 
    e^{ i ( \omega_{i \rightarrow j} - \omega_{j \rightarrow i} )}
    S_i^{-} S_j^{+}
    \\
    -
    ( 1 - \hat{\bd{m}}_i \cdot \hat{\bd{m}}_j )
    e^{ i ( \omega_{i \rightarrow j} + \omega_{j \rightarrow i} )}
    S_i^{-} S_j^{-}
    + \rm{h.c.} \Bigr]
    \,.
    \label{eq:Hbot3}
  \end{multline}
  This expression is manifestly invariant under the local $U(1)$ gauge
  transformation $ \omega_{ i \rightarrow j} \rightarrow \omega_{ i
    \rightarrow j} + \alpha_i$, $S_i^p \rightarrow S_i^p e^{ip
    \alpha_i}$, with $\alpha_i$ arbitrary.  The first term in
  Eq.~(\ref{eq:Hbot3}) shows that a localized spin deviation acquires
  a phase as it moves between sites, which is due to the
  site-dependent orientation of the ground-state magnetization.  The
  local gauge invariance implies conservation of the associated
  current,
  \begin{equation}
    \frac{ \partial \hat{H}^{\bot}   }{ \partial \alpha_i}  
    =  \sum_j 
    \frac{ \partial \hat{H}^{\bot}   }{ \partial \omega_{i \rightarrow j}}  
    =  -  \sum_j J_{ij}  \hat{\bd{m}}_i \cdot \left( {\bd{S}}_i^{\bot}
      \times {\bd{S}}_j^{\bot}  \right)
    \label{eq:currentcon}
    \,.
  \end{equation}
  To see this explicitly, we use the
  equation of motion (\ref{eq:eom}) to write
  \begin{eqnarray}
    \hbar 
    \frac{ \partial {{S}}^{\parallel}_i }{\partial t }
    & = & 
    - \sum_j J_{ij}  \hat{\bd{m}}_i \cdot \left( {\bd{S}}_i^{\bot}
      \times {\bd{S}}_j^{\bot}  \right)
    \nonumber
    \\
    &  & \hspace{-0mm} - {\bd{S}}_i^{\bot} \cdot \Bigl[
    \hat{\bd{m}}_i \times \Big(\,  {\bd{h}}_i - \sum_j J_{ij} S^{\parallel}_j
    \hat{\bd{m}}_j \Big)\, \Bigr]
    \,.
    \label{eq:identity}
  \end{eqnarray}
  Within linear spin-wave theory, the last term may be neglected since 
  it is an order $S^{-1/2}$ smaller than the first term, due to the condition
  (\ref{eq:classical}).
  Taking the thermal average of both sides
  of Eq.~(\ref{eq:identity}) and using the fact that equilibrium
  averages are time-independent, we conclude that the
  average of Eq.~(\ref{eq:currentcon}) indeed vanishes, corresponding
  to a longitudinal spin current with vanishing lattice divergence.
  
  However, in a ring geometry there can be a finite circulating spin
  current in thermal equilibrium provided the classical spin
  configuration ${\bd{m}}_i$ covers a finite total solid angle
  $\Omega$ on the unit sphere in order-parameter space as we move once
  around the ring. In the case of nearest-neighbor coupling, we obtain
  the explicitly gauge-invariant expression $ \Omega = \sum_{i =
    1}^{N} [ \omega_{ i \rightarrow i+1} - \omega_{ i \rightarrow i-1}
  ]$, where $N$ is the number of spins.  Note that $\Omega$ can be
  identified with the total defect angle (``anholonomy'') associated
  with the corresponding parallel transport of a tangential vector
  along a closed path of geodesics \cite{Shapere89}.
  
  We now evaluate the spin current for a ferromagnetic Heisenberg
  chain with nearest-neighbor coupling $J_{i, i \pm 1} = -J<0$.  The
  component of the spin current in the direction $\hat{\bd{m}}_i$, the
  divergence of which appears in Eq.~(\ref{eq:identity}), can be
  written in the gauge invariant form
  \begin{equation}
    I_s = -J \langle \hat{\bd{m}}_i \cdot   \left( {\bd{S}}_i^{\bot}
      \times {\bd{S}}_{i+1}^{\bot}  \right) \rangle = - \partial F_s ( \Omega ) 
    / \partial \Omega
    \,,
    \label{eq:Isres}
  \end{equation}
  where $F_s ( \Omega )$ is the free energy of the spin system.  On
  the other hand, the transverse spin current component in the
  direction of $\hat{\bd{m}}_i \times \hat{\bd{m}}_j$ is an order
  $1/N$ smaller than $I_s$.  In deriving Eq.~(\ref{eq:Isres}), we have
  neglected the terms involving the combinations $S^{-}_iS^{-}_j$ and
  $S^{+}_i S^{+}_j$ in Eq.~(\ref{eq:Hbot3}), because for a ferromagnet
  the contribution of these quantum fluctuations to $I_s$ involves
  higher powers of $1/N$ which are dominant only for $T\to0$.  Since
  the current is a topological property of the system, we may choose
  any convenient geometry for explicit calculations.  For simplicity,
  we shall assume a radial magnetic field ${\bd{B}}_i = |\bd{B}|
  {\bd{r}}_i / | {\bd{r}}_i |$, with the spins located at constant
  latitude $\vartheta_i = \vartheta$ (see Fig.~\ref{fig:angle}).  For $
  | {\bd{h}} | \equiv g \mu_{\text{B}} |\bd{B}| \gtrsim J S (2 \pi / N
  )^2$, the classical ground-state configuration $\hat{\bd{m}}_i$ is
  radial as well, with a slightly different latitude $\vartheta_m$
  satisfying
  \begin{equation}
    \sin ( \vartheta_m - \vartheta ) = - ({JS}/| {\bd{h}} | ) \left[ 
      1 - \cos ( 2 \pi /N  ) \right] \sin ( 2 \vartheta_m )
    \,.
    \label{eq:classicalangle}
  \end{equation} 
  At low temperatures ($T \ll JS$) we may approximate the magnon
  energies by $\epsilon_n+|{\bd{h}}|$ with dispersion $\epsilon_n = JS
  a^2 k_n^2$, where $a$ is the lattice spacing and $k_n = \frac{2 \pi
  }{L} (n - \frac{\Omega }{ 2 \pi} )$ are the quantized wavevectors of
  the magnons on a ring with circumference $L$.  To leading order in
  spin-wave theory, we then obtain the following from
  Eq.~(\ref{eq:Isres}) for the magnetization current $I_m = ( g
  \mu_{\text{B}} / \hbar ) I_s$:
  \begin{equation}
    I_{m} =  \frac{g \mu_{\text{B}}}{ L} \sum_{n}
    \frac{v_n}{e^{ ( \epsilon_n + | {\bd{h}} |  )/ T } -1 }
    \,,
    \label{eq:Issw}
  \end{equation} 
  where $v_n = \hbar^{-1} \partial \epsilon_n / \partial k_n = 2
  JS a^2 k_n / \hbar$.  
  Clearly, Eq.~(\ref{eq:Issw}) is the exact
  bosonic analogue of Eq.~(\ref{eq:Icharge}) \cite{footnoteexact}.
  
  If the temperature is large compared with the level spacing $\Delta
  = JS (2 \pi /N)^2$ between the ground state and the first magnon
  excitation, then Eq.~(\ref{eq:Issw}) can be evaluated analytically,
  \begin{equation}
    I_m = \frac{g \mu_{\text{B}} T}{\hbar} \frac{ \sin \Omega }{ 
      \cos \Omega -\cosh ( 2 \pi \sqrt{ |  {\bd{h}} | / \Delta } )} 
    + O ( e^{ - 2 \pi \sqrt{ \pi T / \Delta } } )
    \label{eq:Imres}
    \,,
  \end{equation}
  which is accurate for $\Delta \ll T \ll JS$. In view of $ | {\bd{h}}
  | / \Delta = S ( g \mu_{\text{B}} |\bd{B}| / JS^2 ) (N / 2 \pi )^2 $,
  it is clear that the persistent spin current is a mesoscopic quantum
  effect, which vanishes for an infinite system or in the classical
  limit $ S \rightarrow \infty$ with constant $JS^2 $.  The radial
  spin configuration shown in Fig.~\ref{fig:angle} yields $\Omega = 2
  \pi ( 1 - \cos \vartheta_m )$ to leading order in $1/N$, so that the
  factor $|\sin \Omega|$ is maximal for $\cos\vartheta_m = 3/4$ or
  $1/4$, i.e., $\vartheta_m \approx 41^{\circ}$ or $76^{\circ}$.  For
  $ | {\bd{h}} | \approx \Delta$ Eq.~(\ref{eq:Imres}) predicts a
  magnetization current on the order of $I_m \approx g \mu_{\text{B}}
  T / \hbar$ around the ring.

\begin{figure}[tb]
    \begin{center}
      \epsfig{file=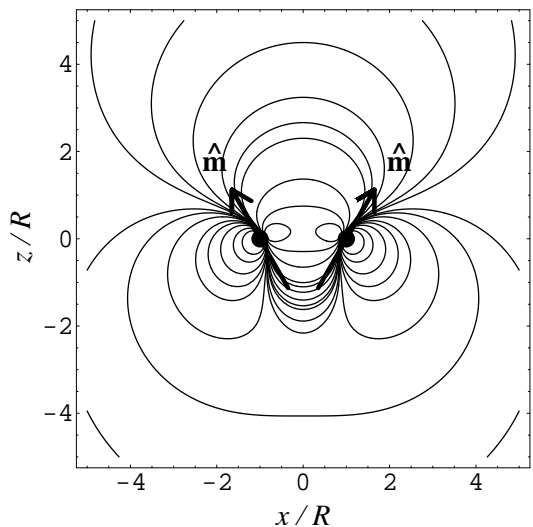,width=70mm}~~~~~~
    \end{center}
    \vspace{-4mm}
    \caption{%
      Lines of constant electric potential
      due to 
      a spin current circulating in a mesoscopic Heisenberg 
      ring with radius $R$ in the geometry of Fig. \ref{fig:angle}
      for $\vartheta_m = 30^{\circ}$.
    }
    \label{fig:field}
  \end{figure}

  Experimentally, the magnetization current can be detected by
  measuring the electric voltage between two points above and below
  the ring. The source of a spatially varying
  magnetization ${\bd{M}} ( {\bd{r}} ) = \sum_{i} \delta ( {\bd{r}} -
  {\bd{r}}_i ) {\bd{m}}_i$ is an effective current density ${\bd{j}} (
  {\bd{r}} ) = c \nabla \times {\bd{M}} ( {\bd{r}} )$.  If the
  magnetization is moving with velocity $ {\bd{v}} ( {\bd{r}}) $, it
  is accompanied by a polarization ${\bd{P}} ( {\bd{r}} ) = \frac{
    {\bd{v}} ( {\bd{r}} )}{c} \times {\bd{M}} ( {\bd{r}} )$ to leading
  order in $ {\bd{v}} ( {\bd{r}}) /c$.  This is easily shown by means
  of a Lorentz boost to the rest frame of the magnetic dipoles
  \cite{Hirsch99}.  The polarization corresponds to a charge density
  $\rho ( {\bd{r}} ) = - \nabla \cdot {\bd{P}} ( {\bd{r}}) $, which in
  turn generates an electric field ${\bd{E}} ( {\bd{r}} ) = - \nabla
  \phi ( {\bd{r}} )$, with the scalar potential $ \phi ( {\bd{r}} ) =
  \int d^3 r^{\prime} \rho ( {\bd{r}}^{\prime} ) / | {\bd{r}} -
  {\bd{r}}^{\prime} | $.  Combining the above relations, we obtain a
  generalized Biot-Savart law for the scalar potential due to magnetic
  dipole currents,
  \begin{eqnarray}
    \phi ( {\bd{r}} )  & = & \frac{1}{c}  \int d^3 r^{\prime} 
    [ {\bd{v}} ( {\bd{r}}^{\prime} ) \times {\bd{M}} ( {\bd{r}}^{\prime} ) ]
    \cdot \frac{  ( {\bd{r}} - {\bd{r}}^{\prime} ) }{ 
      | {\bd{r}} - {\bd{r}}^{\prime} |^3 }
    \nonumber
    \\
    & = & 
    \frac{ I_m}{c} \oint
    [ d {\bd{r}}^{\prime} \times \hat{\bd{m}} ( {\bd{r}}^{\prime} )  ] \cdot
    \frac{ {\bd{r}} - {\bd{r}}^{\prime} }{ | {\bd{r}} - {\bd{r}}^{\prime} |^3 }
    \label{eq:Biotsavart}
    \,.
  \end{eqnarray}
  The second line is valid for a current loop, where $ {\bd{v}} |
  {\bd{M}} | d^{3}r = I_m d {\bd{r}}$.  Although
  Eq.~(\ref{eq:Biotsavart}) is of fundamental importance in spin
  transport, we have not been able to find it in standard texts on
  classical electrodynamics.
  The expression derived in Ref.~\onlinecite{Meier02} for a straight
  line is a special case of Eq.~(\ref{eq:Biotsavart}).  For the simple
  geometry shown in Fig.~\ref{fig:angle} the integration in
  Eq.~(\ref{eq:Biotsavart}) can be reduced to elliptic integrals. The
  equipotentials are shown in Fig.~\ref{fig:field}.
  In the far zone, the electric field approaches a dipole field, with
  potential $ \phi ( {\bd{r}} ) = {\bd{p}} \cdot {\bd{r}} / | {\bd{r}}
  |^3$ and dipole moment ${\bd{p}} = - {\bd{e}}_z ( I_m / c) L \sin
  \vartheta_m $.  To estimate the order of magnitude of the
  magnetization current, consider a mesoscopic $S=1/2$ Heisenberg
  chain with $g=2$, $N = 100$, and $J = 100\text{K}$.  Then the
  condition $g \mu_{\text{B}} |\bd{B}| \approx \Delta$ is satisfied
  for $|\bd{B}| \approx 0.1\text{T}$.  To obtain a sizable $\Omega$,
  one should generate inhomogeneities of the magnetic field in the
  submicron range; these may be achievable in the vicinity of a
  magnetic flux line trapped in a type-II superconductor.  In the
  dipole approximation, the potential drop between two points located a
  distance $d$ above and below the loop on the $z$-axis is given by $
  U \approx
  0.5\text{nV}\cdot(T/\text{K})(L/\text{nm})/(d/\text{nm})^2$.  For $T
  = 60\text{K}$ and $d=L=100\text{nm}$ this yields a voltage $U
  \approx 0.3\text{nV}$.  However, its experimental detection is
  difficult, because mobile charges will tend to screen this static
  dipole field.  Note that the magnetization current in
  Eq.~(\ref{eq:Issw}) involves only a single factor of $g
  \mu_{\text{B}}$, whereas for a two-terminal geometry the current is
  proportional to $( g \mu_{\text{B}} )^2$ \cite{Meier02}.
  
  In summary, we have shown that a ferromagnetic Heisenberg ring in an
  inhomogeneous magnetic field can support a persistent magnetization
  current, which is the precise bosonic analogue of the persistent
  charge current in normal metal rings.  The magnetization current is
  a mesoscopic quantum interference effect and flows without
  dissipation, corresponding to an infinite spin conductance.  For
  weak magnetic fields ($ g \mu_{\text{B}} |\bd{B}| \approx \Delta$)
  the magnetization current gives rise to an electric dipole field.
  For larger fields, the ballistic contribution to the spin current
  considered in this work is exponentially suppressed.  By analogy
  with the persistent charge current in normal metal rings, we expect
  that in this regime the persistent magnetization current is
  dominated by collective phenomena such as spin diffusion and weak
  localization effects.  The calculation of the dominant contribution
  in this regime is still an open problem.  With suitable magnetic
  fields, persistent magnetization currents should also exist in
  antiferromagnetic or ferrimagnetic spin chains, where we expect
  finite currents at $T=0$ due to quantum fluctuations.

  This work was supported by the DFG via Forschergruppe FOR 412,
  Project No. KO 1442/5-1.



\begin{thebibliography}{99}
\bibitem{footnoteexact}%
    To compare Eqs.~(\ref{eq:Icharge}) and (\ref{eq:Issw}), we should
    divide the currents by the corresponding ``charges'', i.e., the
    particle number current $I / (-e)$ should be compared with $I_s /
    \hbar = I_m/ ( g \mu_{\text{B}} )$.
\end{thebibliography}
\end{document}